# Ternary-composition tuning of the anomalous Nernst effect in amorphous-like Gd–Co–Pt films

Tomohiro Koizumi(小泉朋寛)[1], Mojtaba Mohammadi[1], Hiroto Imaeda(今枝寛人)[1], Hiroyuki Awano(粟野博之)[1], and Kenji Tanabe(田辺賢士)[1*]
[1]Toyota Technological Institute, Nagoya, 468-8511, Japan
*electronic mail: tanabe@toyota-ti.ac.jp

## Abstract

We report composition tuning of the anomalous Nernst effect in amorphous-like Gd–Co–Pt ternary films. Pt incorporation into Gd–Co films modifies the anomalous Nernst coefficient and induces a sign reversal of $S_{\mathrm{ANE}}$, whereas Gd incorporation into Co–Pt films suppresses $S_{\mathrm{ANE}}$ but reduces thermal conductivity. Owing to the balance between transverse thermoelectric response and thermal transport, the heat-flux sensitivity reaches approximately 0.24 μm/A. Composition maps reveal that the magnitude and heat-flux sensitivity of the anomalous Nernst effect can be systematically tuned in the Gd–Co–Pt ternary composition space. This work extends ANE material design from binary-alloy optimization to ternary-composition engineering.

The anomalous Nernst effect (ANE) has attracted considerable attention as a transverse thermoelectric phenomenon in magnetic materials [1-14]. In the ANE, an electric field is generated perpendicular to both the magnetization and the heat flux, enabling a planar device geometry in which the output voltage can be extracted along the film plane. This configuration is particularly attractive for thin-film heat-flux sensors (HFSs) because thermoelectric elements can be connected laterally without stacking many junctions along the temperature-gradient direction [15-23]. Therefore, magnetic thin films that combine a large transverse thermoelectric response with low thermal conductivity are desirable for realizing sensitive and scalable HFSs.

Co–Pt alloys are promising materials for transverse thermoelectric conversion because Pt provides strong spin–orbit coupling and modifies spin–orbit-coupled transport in Co-based ferromagnets [24-26]. Recent studies on Co–Pt thin films have shown that composition tuning can strongly enhance the anomalous Nernst coefficient and anomalous Nernst conductivity, even in chemically disordered or partially crystalline films. In particular, disordered Co–Pt films exhibit a large anomalous Nernst coefficient near the equiatomic composition, together with a high transverse thermoelectric conductivity and sizable heat-flux sensitivity [26]. These results indicate that Co–Pt-based alloys provide a useful platform for achieving large transverse thermoelectric responses without requiring long-range chemical ordering.

However, further material design is required for heat-flux sensing applications. The heat-flux sensitivity is governed not only by the anomalous Nernst coefficient but also by the thermal conductivity. Thus, reducing thermal conductivity while maintaining the transverse thermoelectric response is an important strategy for increasing the electric-field response per heat flux density. Since the heat-flux sensitivity is approximately proportional to $S_{\mathrm{ANE}}/\kappa$, materials that simultaneously exhibit a large anomalous Nernst coefficient and low thermal conductivity are particularly attractive for ANE-based heat-flux sensing. Rare-earth–transition-metal amorphous ferrimagnetic alloys, such as Gd–Co, are attractive in this respect because they possess low thermal conductivity, ferrimagnetic compensation behavior, and composition-tunable ANE [27-36]. In addition, the incorporation of a rare-earth element can provide an additional degree of freedom for controlling the sign and magnitude of the ANE through ferrimagnetic coupling.

In this study, we investigate amorphous-like Gd–Co–Pt ternary thin films as an extension of Co–Pt and Gd–Co alloy platforms. The concept is to combine the large spin–orbit-coupled transverse thermoelectric response of Co–Pt with the low-thermal-conductivity and ferrimagnetic tunability of Gd–Co. We show that Pt incorporation into Gd–Co films modifies the anomalous Nernst coefficient and induces a sign reversal of $S_{\mathrm{ANE}}$, suggesting that Pt contributes to the transverse thermoelectric response through Co–Pt-like spin–orbit-coupled electronic states. Conversely, Gd incorporation into Co–Pt films strongly suppresses $S_{\mathrm{ANE}}$, but the heat-flux sensitivity $E/j_{\mathrm{Q}}$ is less severely reduced because of the concurrent reduction in thermal conductivity. By mapping the ANE response in the Gd–Co–Pt ternary composition space, we demonstrate that the magnitude and heat-flux sensitivity of the ANE can be systematically tuned. These results suggest that Gd–Co–Pt alloys provide a ternary-composition platform for controlling transverse thermoelectric transport and thermal transport in ANE-based heat-flux sensing materials.

Amorphous-like Gd–Co–Pt thin films were deposited on $SiO_2$ glass substrates by magnetron sputtering at room temperature. The film structure was $Si_3N_4$ (3 nm)/Gd–Co–Pt (20 nm)/$Si_3N_4$ (10 nm). The alloy composition was controlled by changing the sputtering powers of Gd, Co, and Pt targets and was determined by energy-dispersive X-ray spectroscopy. The main composition series were $(Gd_{15}Co_{85})_{100-x}Pt_x$, $(Co_{50}Pt_{50})_{100-x}Gd_x$, and $(Gd_{25}Co_{75})_{100-x}Pt_x$. The structural properties of the Gd–Co–Pt films were examined by X-ray diffraction. Most Gd–Co–Pt films exhibited amorphous-like X-ray diffraction patterns without distinct crystalline reflections. Weak fcc (111)-type diffraction peaks were observed only in Co–Pt-rich compositions containing 5 at.% Gd or less (Fig. 4(d)).

The ANE was measured using a Hall-bar-shaped sample on $SiO_2$ glass substrates as shown in Fig. 1(a). A temperature gradient was applied along $y$ direction, and the transverse voltage was measured along $x$ direction under an external magnetic field. The anomalous Nernst coefficient $S_{\mathrm{ANE}}$ was obtained from the slope of the electric field $E$ as a function of the temperature gradient $\nabla T$. Longitudinal resistivity, anomalous Hall resistivity, and Seebeck coefficient were measured independently to evaluate the anomalous Nernst conductivity using $S_{\mathrm{ANE}} = \rho_{xx}\alpha_{xy} + \rho_{xy}\alpha_{xx}$, where $\alpha_{xx}$ is derived by $\alpha_{xx} = S_{xx}/\rho_{xx}$. The heat-flux sensitivity $E/j_{\mathrm{Q}}$ was determined by applying a controlled heat flux density $j_{\mathrm{Q}}$ through the film and measuring the resulting ANE electric field [37-38]. The heat flux density was calibrated using a HFS placed in the heat path. Thermal conductivity was derived by using values of $S_{\mathrm{ANE}}$ and sensitivity $E/j_{\mathrm{Q}}$, following the previous report [20]. Detailed measurement configurations are shown schematically in Figs. 1 and 3.

Figure 1 shows the ANE in amorphous $(Gd_{15}Co_{85})_{100-x}Pt_x$ thin films. The films exhibited clear magnetic-field-dependent ANE voltages, as shown in Fig. 1(b), confirming that a transverse thermoelectric signal was generated in the Gd–Co–Pt films.

The electric field was proportional to the applied temperature gradient (Fig. 1(c)), indicating that the anomalous Nernst coefficient $S_{\mathrm{ANE}}$ was well defined in the present measurement configuration.

Figure 1(d) summarizes the Pt-content dependence of $S_{\mathrm{ANE}}$ in $(Gd_{15}Co_{85})_{100-x}Pt_x$ films. $S_{\mathrm{ANE}}$ increases nearly monotonically with Pt content, reaching approximately 2 μV/K in the investigated composition range. This value is roughly twice that typically reported for binary Gd–Co and Tb–Co ferrimagnetic alloys, indicating that Pt incorporation substantially enhances the transverse thermoelectric response while preserving the amorphous-like ferrimagnetic character of the films. Because Co–Pt alloys are known to exhibit large spin–orbit-coupled transverse thermoelectric transport, the present result suggests that the Co–Pt-derived contribution is partially retained even when the alloy is converted into a Gd-containing ferrimagnetic ternary system. Since the investigated compositions remain predominantly amorphous-like, the enhancement of $S_{\mathrm{ANE}}$ cannot be attributed primarily to long-range crystallographic ordering and instead reflects changes in electronic transport induced by Pt incorporation.

To clarify the origin of the Pt-content dependence, we analyzed the electrical and thermoelectric transport coefficients, as summarized in Fig. 2. The longitudinal resistivity $\rho_{xx}$, anomalous Hall resistivity $\rho_{xy}$, anomalous Hall angle, and longitudinal Seebeck coefficient $S_{xx}$ vary systematically with Pt content. The anomalous Nernst coefficient was decomposed into $S_1 = \rho_{xx}\alpha_{xy}$ and $S_2 = \rho_{xy}\alpha_{xx}$. As shown in Fig. 2, $S_{\mathrm{ANE}}$ approximately follows the trend of $S_1$, whereas $S_2$ remains relatively small over the investigated Pt-content range. This indicates that the Pt-induced variation in $S_{\mathrm{ANE}}$ is mainly associated with the transverse thermoelectric conductivity term rather than with the anomalous-Hall-related contribution.

The anomalous Nernst conductivity $\alpha_{xy}$ increases monotonically from approximately 0.5 to 1.7 A $m^{-1}$ $K^{-1}$ with increasing Pt content. This behavior is consistent with the role of Pt in modifying spin–orbit-coupled transverse transport in Co-based alloys. However, the increase in $\alpha_{xy}$ in Gd–Co–Pt is smaller than that in binary Co–Pt films, suggesting that the addition of Gd reduces the Co–Pt-derived transverse thermoelectric response while introducing additional advantages such as ferrimagnetic tunability and reduced thermal conductivity. Thus, the role of Pt in the ternary alloy is not to produce a dramatic enhancement of $S_{\mathrm{ANE}}$, but to provide a controllable transverse thermoelectric contribution within a low-thermal-conductivity ferrimagnetic matrix.

We next evaluated the heat-flux sensitivity of the Gd–Co–Pt films. Figure 3(a) shows the measurement configuration, in which a heat flux density $j_{\mathrm{Q}}$ was applied through the sample and the resulting ANE electric field was detected along the in-plane direction. The ANE voltage exhibited a clear magnetic-field-dependent response under applied heat flux density, as shown in Fig. 3(b). The electric field $E$ increased linearly with $j_{\mathrm{Q}}$, confirming that the heat-flux response was in the linear regime in Fig. 3(c). The heat-flux sensitivity was then defined as $E/j_{\mathrm{Q}}$, corresponding to the electric field generated per unit heat flux density.

Figure 3(d) summarizes the Pt-content dependence of $E/j_{\mathrm{Q}}$ in $(Gd_{15}Co_{85})_{100-x}Pt_x$ films. The sensitivity varies systematically with Pt content and reaches approximately 0.24 μm/A. This value is slightly lower than the maximum value reported for disordered Co–Pt films [26], but it is comparable to those of amorphous Gd–Co alloys [20]. This result indicates that Gd–Co–Pt films retain a relatively high heat-flux sensitivity despite the incorporation of Gd into the Co–Pt-based alloy system. Importantly, the purpose of Gd incorporation is not simply to maximize $S_{\mathrm{ANE}}$ or $\alpha_{xy}$, but to introduce additional material-design parameters, including ferrimagnetic tunability, sign control, and thermal-conductivity reduction.

Thermal conductivity is an essential parameter for ANE-based heat-flux sensing because the heat-flux sensitivity is approximately governed by the ratio between the anomalous Nernst coefficient and the thermal conductivity. Figure 3(e) shows the Pt-content dependence of the thermal conductivity in $(Gd_{15}Co_{85})_{100-x}Pt_x$ films, and it can be seen that even with Pt doping, the thermal conductivity is kept sufficiently low below 10 W/mK due to the presence of Gd. Such low values are considerably smaller than those of conventional crystalline metallic alloys and reflect the amorphous-like nature of the Gd–Co–Pt films. The preservation of low thermal conductivity is crucial for maintaining high heat-flux sensitivity despite the introduction of Pt.

To further clarify the distinct contributions of Pt and Gd to transverse thermoelectric transport and thermal transport, we next investigate two complementary composition series in Figure 4. In the $(Co_{50}Pt_{50})_{100-x}Gd_x$ series, $S_{\mathrm{ANE}}$ decreases rapidly with increasing Gd content in Fig. 4(a), indicating that the Co–Pt-derived anomalous Nernst response is strongly suppressed by Gd incorporation. However, the heat-flux sensitivity $E/j_{\mathrm{Q}}$ does not decrease as rapidly as $S_{\mathrm{ANE}}$ in Fig. 4(b). This difference suggests that Gd incorporation simultaneously reduces the thermal conductivity as shown in Fig. 4(c), thereby partially compensating for the reduction in anomalous Nernst thermopower. Indeed, the X-ray diffraction patterns in Fig. 4(d) show that Gd doping disrupts the crystallinity, which may contribute to the suppression of the thermal conductivity. These results demonstrate the complementary roles of Pt and Gd: Pt mainly tunes the transverse thermoelectric response, whereas Gd provides ferrimagnetic tunability and thermal-conductivity reduction.

In contrast, in the $(Gd_{25}Co_{75})_{100-x}Pt_x$ series, $S_{\mathrm{ANE}}$ changes systematically with Pt content and even reverses its sign as shown in Fig. 4(e). This sign reversal is difficult to explain if Pt acts only as a nonmagnetic diluent. Instead, it suggests that Pt modifies the Co-derived transverse thermoelectric response through Co–Pt hybridization and spin–orbit-coupled electronic states. A possible contribution from proximity-induced Pt spin polarization may also be relevant, although element-specific magnetic measurements would be required to clarify this point.

The ternary composition maps in Figures 4(g and h) provide a comprehensive overview of the compositional dependence of transverse thermoelectric and heat-flux-sensing performance. The $S_{\mathrm{ANE}}$ map exhibits a clear reduction of the magnitude in the Gd–Co–Pt composition space by the Gd-doping, while the $E/j_{\mathrm{Q}}$ map shows that high heat-flux sensitivity is maintained over a broader composition region than expected from the $S_{\mathrm{ANE}}$ map alone. Therefore, the Gd–Co–Pt ternary system provides a composition-design platform in which the magnitude and heat-flux sensitivity of the anomalous Nernst response can be tuned independently to some extent.

In summary, we demonstrated ternary-composition tuning of the anomalous Nernst effect in amorphous-like Gd–Co–Pt thin films. Pt incorporation into Gd–Co films induced a sign reversal of $S_{\mathrm{ANE}}$, suggesting a Co–Pt-like contribution to the transverse thermoelectric response. In contrast, Gd incorporation into Co–Pt films suppressed $S_{\mathrm{ANE}}$ but reduced the thermal conductivity, thereby maintaining a relatively high heat-flux sensitivity. The maximum $E/j_{\mathrm{Q}}$ reached approximately 0.24 μm/A. The ternary composition maps show that the magnitude and heat-flux sensitivity of the ANE can be systematically tuned in the Gd–Co–Pt composition space. This work extends the material design of ANE-based heat-flux sensors from binary-alloy optimization to ternary-composition engineering.

## ACKNOWLEDGEMENTS

This work was partly supported by the Research Center for Smart Photons and Materials in Toyota Technological Institute, the Fuji Science and Technology Foundation and JST A-STEP (Grant No. JPMJTR25TA).

## AUTHOR DECLARATIONS

### Conflict of Interest

The authors have no conflicts to disclose.

## Author Contributions

**Tomohiro Koizumi**: Data Curation (lead); Investigation (lead); Formal Analysis (lead); Writing/Original Draft Preparation (supporting); Writing/Review & Editing (supporting); **Mojtaba Mohammadi**: Investigation (supporting); Formal Analysis (supporting); Writing/Review & Editing (supporting); **Hiroto Imaeda**: Investigation (supporting); Formal Analysis (supporting); Writing/Review & Editing (supporting); **Hiroyuki Awano**: Investigation (supporting); Resources (equal); Writing/Review & Editing (supporting); **Kenji Tanabe**: Conceptualization (lead); Data Curation (supporting); Formal Analysis (supporting); Funding Acquisition (lead); Investigation (supporting); Methodology (lead); Project Administration (lead); Resources (equal); Supervision (lead); Writing/Original Draft Preparation (lead); Writing/Review & Editing (lead)

## DATA AVAILABILITY

The data that support the findings of this study are available from the corresponding author upon reasonable request.

## Figure Caption

Figure 1. (a) Schematic illustration of the measurement geometry for the ANE. An in-plane temperature gradient, $(-\nabla T)$, is applied along the $y$ direction, and the anomalous Nernst electric field, $E$, is measured along the $x$ direction under an out-of-plane magnetic field, $H$. (b) Anomalous Nernst voltage as a function of the magnetic field. (c) Anomalous Nernst electric field as a function of the temperature gradient for $(Gd_{15}Co_{85})_{100-x}Pt_x$ with ($x$ = 55.4). The dashed line represents a linear fit. (d) Pt-content dependence of the anomalous Nernst coefficient, $S_{\mathrm{ANE}}$, for $(Gd_{15}Co_{85})_{100-x}Pt_x$ thin films.

Figure 2. (a) Longitudinal resistivity, $\rho_{xx}$, (b) anomalous Hall resistivity, $\rho_{xy}$, (c) anomalous Hall angle, $|\rho_{xy}/\rho_{xx}|$, (d) longitudinal thermopower, $S_{xx}$, (e) anomalous Nernst coefficient, $S_{\mathrm{ANE}}$, and its decomposed contributions $S_1$ and $S_2$, and (f) anomalous Nernst conductivity, $\alpha_{xy}$, as functions of Pt content, for $(Gd_{15}Co_{85})_{100-x}Pt_x$ thin films.

Figure 3. (a) Schematic diagram of measurement geometry for the sensitivity $E/j_{\mathrm{Q}}$. A heat-flux density $j_{\mathrm{Q}}$ is applied along the $z$ direction, and the anomalous Nernst electric field, $E$, is measured along the $x$ direction under an in-plane magnetic field, $H$. (b) Anomalous Nernst voltage as a function of the magnetic field. (c) Anomalous Nernst electric field as a function of the heat-flux density for $(Gd_{15}Co_{85})_{100-x}Pt_x$ with ($x$ = 55.4). The dashed line represents a linear fit. (d) Pt-content dependence of the heat-flux sensitivity, $E/j_{\mathrm{Q}}$, for $(Gd_{15}Co_{85})_{100-x}Pt_x$ thin films. (e) Thermal conductivity as a function of Pt content for $(Gd_{15}Co_{85})_{100-x}Pt_x$.

Figure 4. (a) $S_{\mathrm{ANE}}$ for $(Co_{50}Pt_{50})_{100-x}Gd_x$. (b) Heat-flux sensitivity $E/j_{\mathrm{Q}}$ for $(Co_{50}Pt_{50})_{100-x}Gd_x$. (c) Thermal conductivity $\kappa$ for $(Co_{50}Pt_{50})_{100-x}Gd_x$. (d) X-ray diffraction patterns for $(Co_{50}Pt_{50})_{100-x}Gd_x$. (e) $S_{\mathrm{ANE}}$ for $(Gd_{25}Co_{75})_{100-x}Pt_x$. (f) Heat-flux sensitivity $E/j_{\mathrm{Q}}$ for $(Gd_{25}Co_{75})_{100-x}Pt_x$. (g) Ternary composition map of $S_{\mathrm{ANE}}$. (h) Ternary composition map of $E/j_{\mathrm{Q}}$. The results show that Gd incorporation into Co–Pt suppresses $S_{\mathrm{ANE}}$ but preserves relatively high $E/j_{\mathrm{Q}}$ through a concurrent reduction in thermal conductivity.

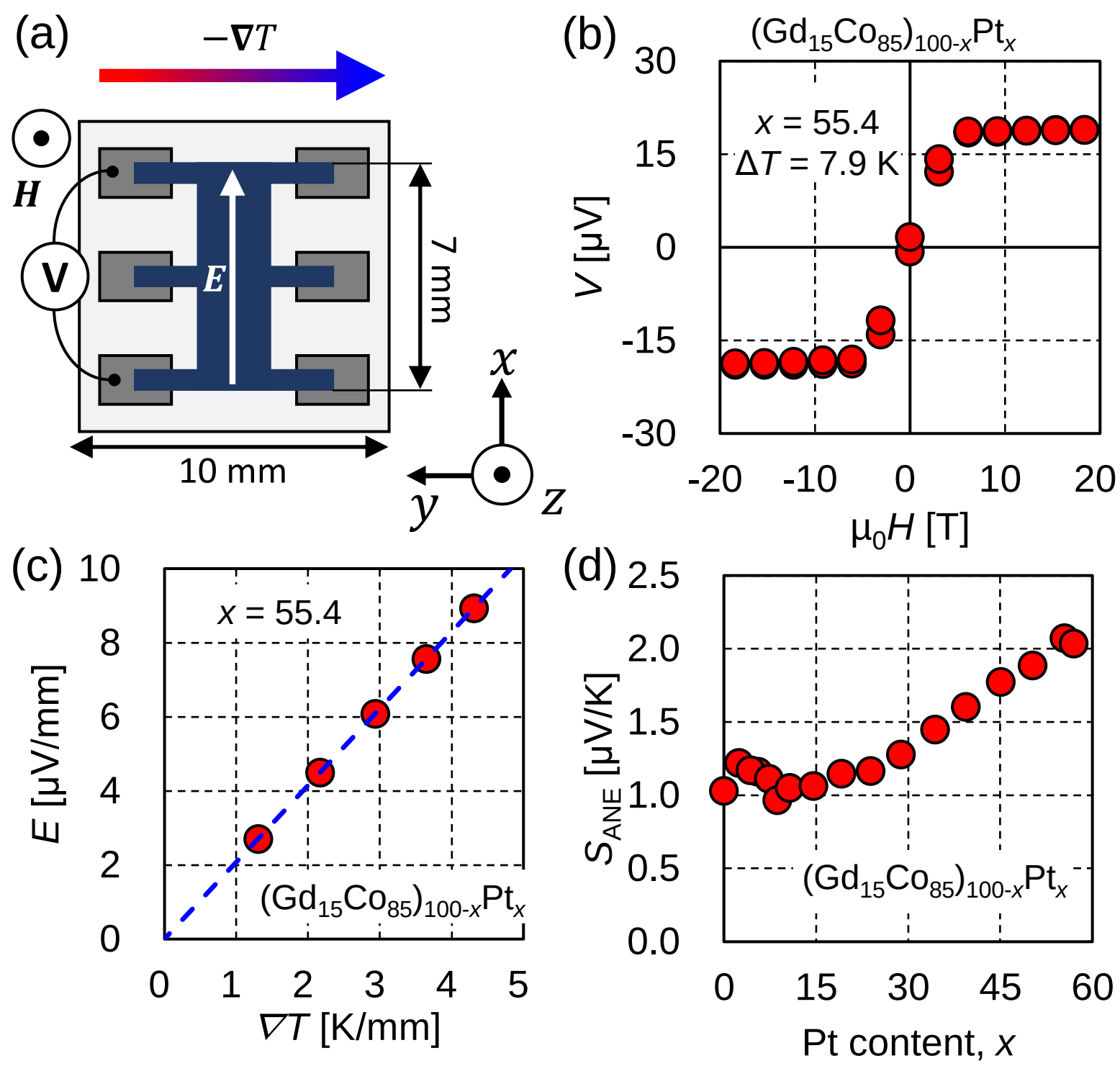
(a)
$-\nabla T$
H
V
E
7 mm
10 mm
x
y
z
(b)
$(Gd_{15}Co_{85})_{100-x}Pt_x$
x = 55.4
ΔT = 7.9 K
V [μV]
$\mu_0 H$ [T]
(c)
x = 55.4
$(Gd_{15}Co_{85})_{100-x}Pt_x$
E [μV/mm]
∇T [K/mm]
(d)
$(Gd_{15}Co_{85})_{100-x}Pt_x$
$S_{ANE}$ [μV/K]
Pt content, x

Fig. 1

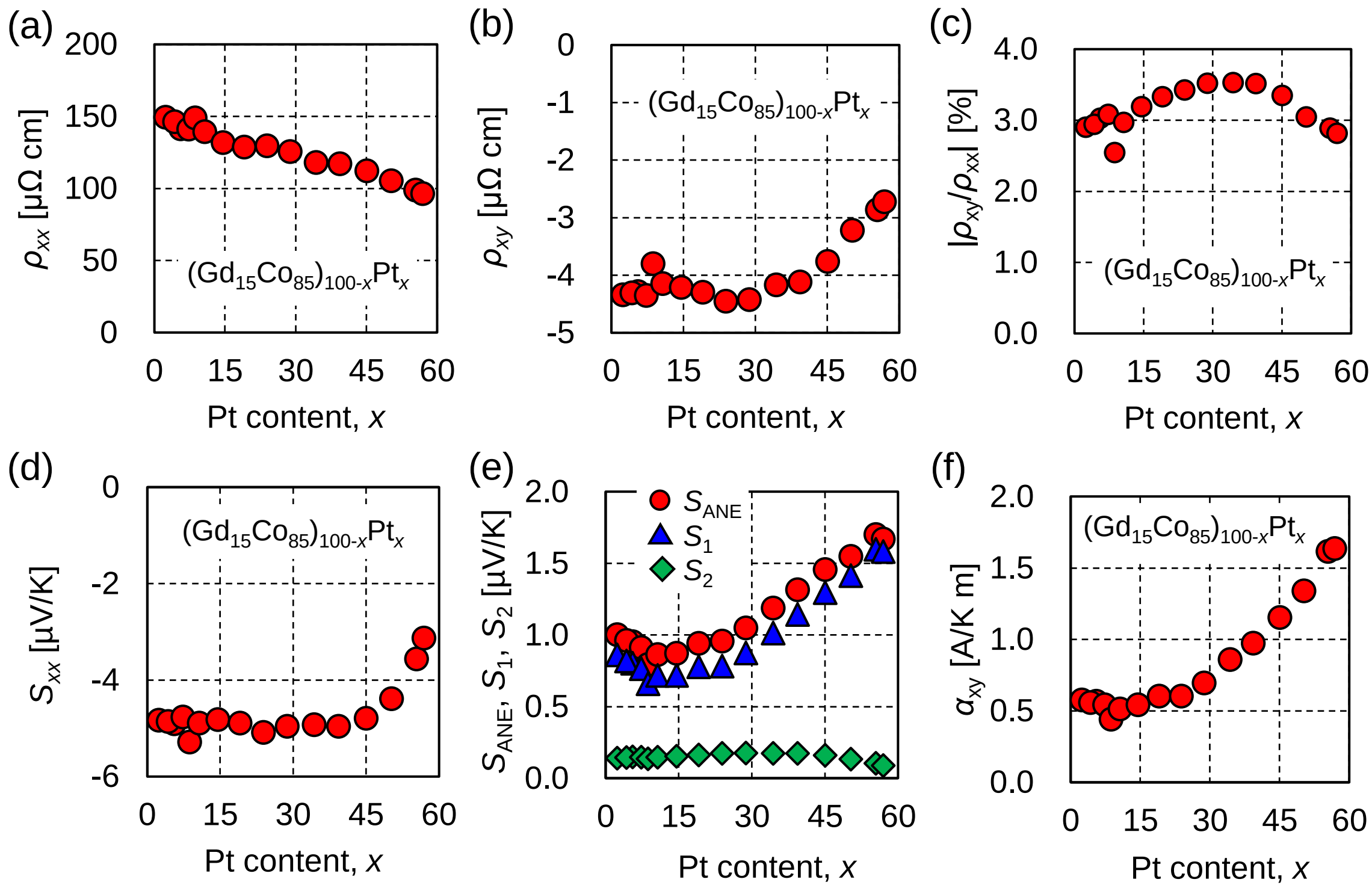

(a)
$\rho_{xx}$ [μΩ cm]
$(Gd_{15}Co_{85})_{100-x}Pt_x$
Pt content, $x$
(b)
$\rho_{xy}$ [μΩ cm]
$(Gd_{15}Co_{85})_{100-x}Pt_x$
Pt content, $x$
(c)
$|\rho_{xy}/\rho_{xx}|$ [%]
$(Gd_{15}Co_{85})_{100-x}Pt_x$
Pt content, $x$
(d)
$S_{xx}$ [μV/K]
$(Gd_{15}Co_{85})_{100-x}Pt_x$
Pt content, $x$
(e)
$S_{ANE}$, $S_1$, $S_2$ [μV/K]
$S_{ANE}$
$S_1$
$S_2$
Pt content, $x$
(f)
$\alpha_{xy}$ [A/K m]
$(Gd_{15}Co_{85})_{100-x}Pt_x$
Pt content, $x$

Fig. 2

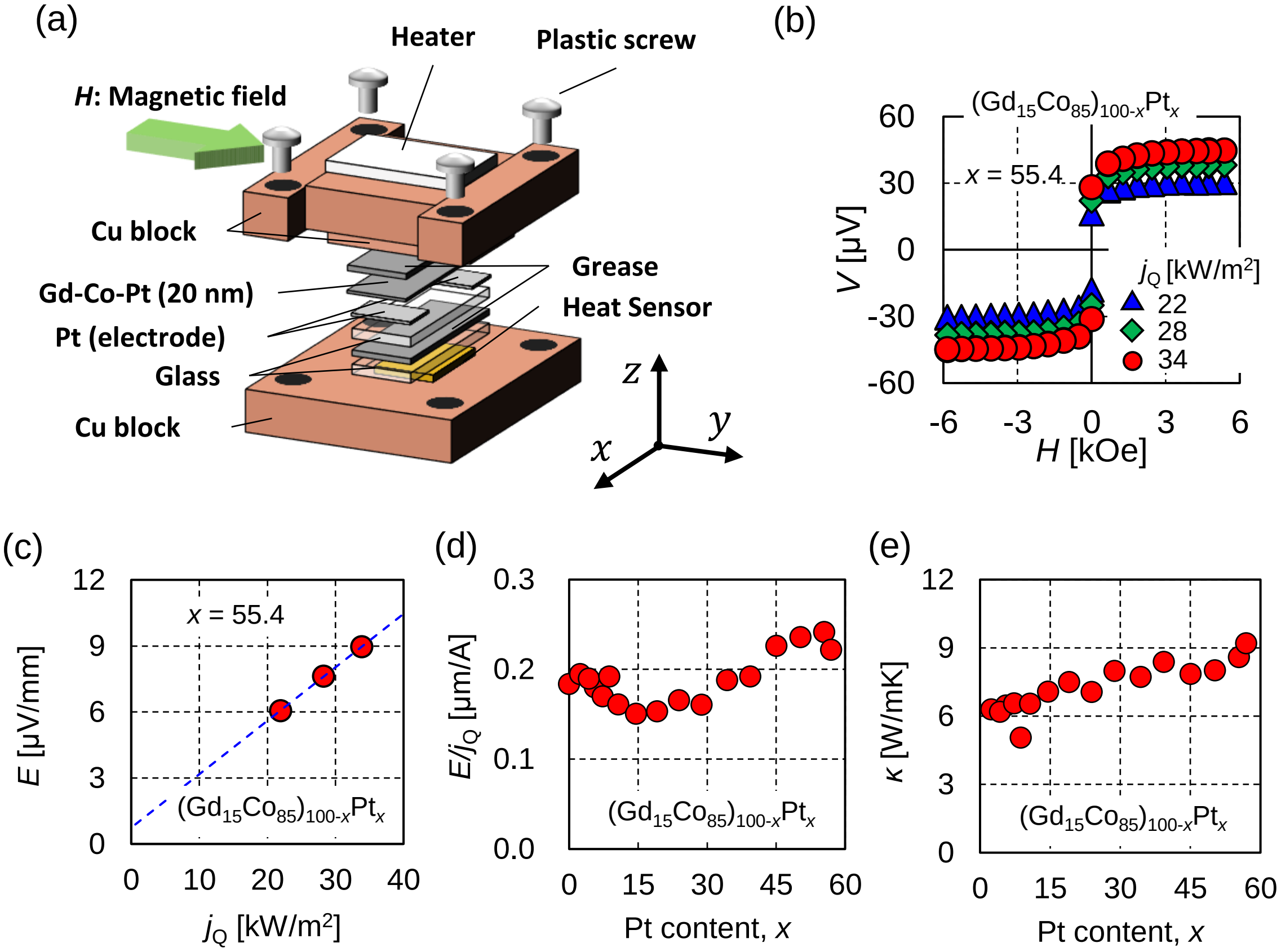

(a)
Heater
Plastic screw
H: Magnetic field
Cu block
Gd-Co-Pt (20 nm)
Pt (electrode)
Glass
Cu block
Grease
Heat Sensor
z
y
x
(b)
(Gd15Co85)100-xPtx
x = 55.4
V [μV]
H [kOe]
jQ [kW/m2]
22
28
34
(c)
x = 55.4
E [μV/mm]
(Gd15Co85)100-xPtx
jQ [kW/m2]
(d)
E/jQ [μm/A]
(Gd15Co85)100-xPtx
Pt content, x
(e)
κ [W/mK]
(Gd15Co85)100-xPtx
Pt content, x

Fig. 3

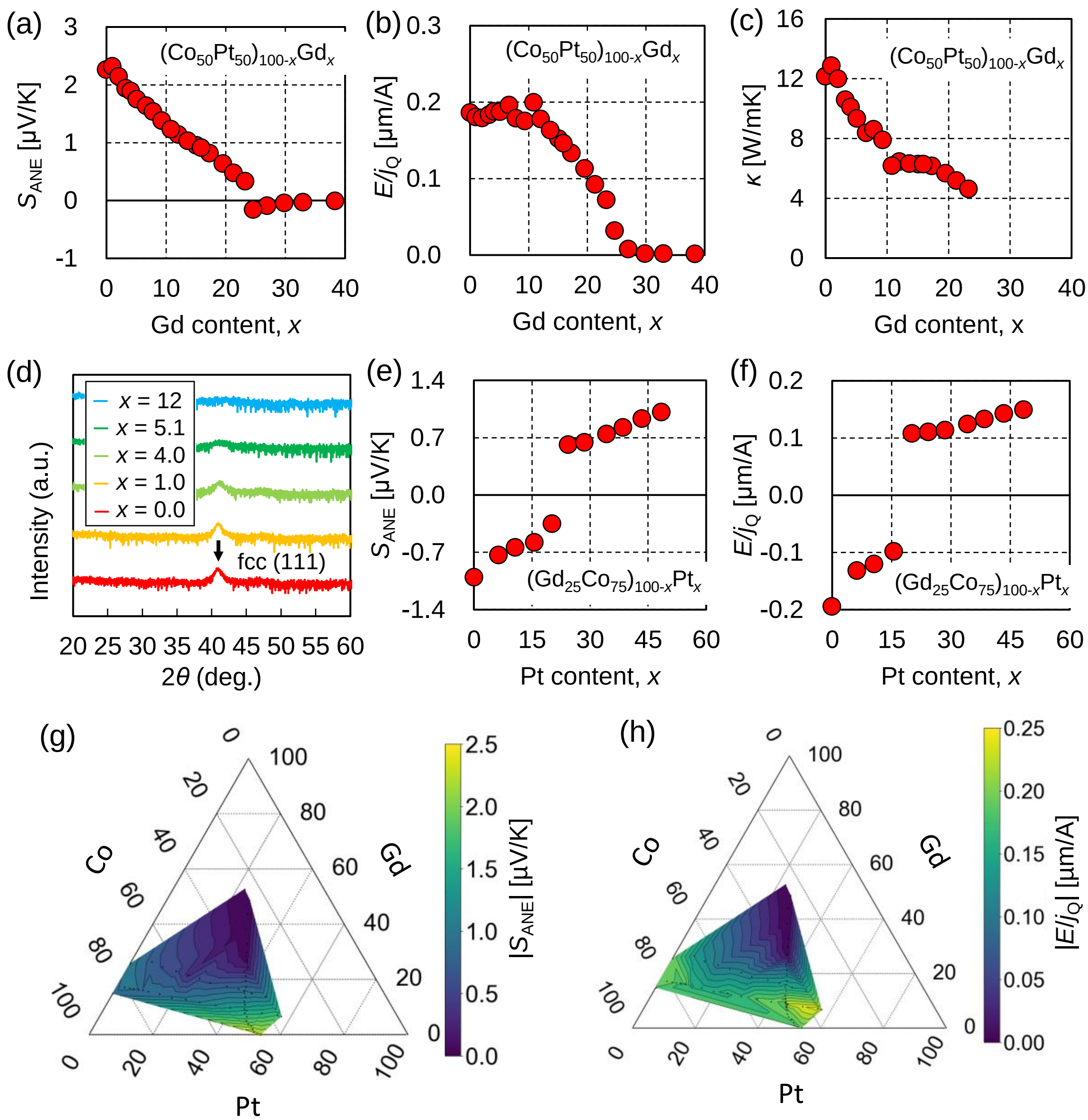

(a)
$(Co_{50}Pt_{50})_{100-x}Gd_x$
$S_{ANE}$ [μV/K]
Gd content, $x$
(b)
$(Co_{50}Pt_{50})_{100-x}Gd_x$
$E/j_Q$ [μm/A]
Gd content, $x$
(c)
$(Co_{50}Pt_{50})_{100-x}Gd_x$
$\kappa$ [W/mK]
Gd content, x
(d)
$x$ = 12
$x$ = 5.1
$x$ = 4.0
$x$ = 1.0
$x$ = 0.0
fcc (111)
Intensity (a.u.)
$2\theta$ (deg.)
(e)
$(Gd_{25}Co_{75})_{100-x}Pt_x$
$S_{ANE}$ [μV/K]
Pt content, $x$
(f)
$(Gd_{25}Co_{75})_{100-x}Pt_x$
$E/j_Q$ [μm/A]
Pt content, $x$
(g)
Co
Gd
Pt
$|S_{ANE}|$ [μV/K]
(h)
Co
Gd
Pt
$|E/j_Q|$ [μm/A]

Fig. 4